\documentclass[12pt]{article}
\usepackage{draft1,tikz-cd}
\usepackage[mathscr]{eucal}

\def\ma{\mathcal}

\begin{document}

\begin{titlepage}

\begin{center}

\hfill \\
\hfill \\
\vskip 1cm

\title{Celestial self-dual Yang-Mills theory: a new formula and the OPE limit}

\bigskip

\author{Pratik Chattopadhyay$^{a}$ and Yi-Xiao Tao$^{b}$}

\bigskip

\address{\small${}^a$Namtech, IIT Gandhinagar campus, Gandhinagar 382355, India\\
\small${}^b$ Yau Mathematical Sciences Center (YMSC), Tsinghua University, Beijing 100084, China}
\email{pratikpc@gmail.com, taoyx21@mails.tsinghua.edu.cn}

\end{center}

\vfill

\begin{abstract}
Celestial holography is a new way to understand flat-space amplitudes. Self-dual theories, due to their nice properties, are good subjects to study celestial holography. In this paper, we developed a new formula to calculate the celestial color-ordered self-dual Yang-Mills amplitudes based on celestial Berends-Giele currents, which makes the leading OPE limit manifest. In addition, we explore some higher-order terms of OPE in the celestial self-dual Yang-Mills theory.

\end{abstract}

\vfill

\end{titlepage}

\tableofcontents

\newpage

\section{Introduction}
Inspired by AdS/CFT \cite{Maldacena:1997re}, people want to find similar correspondence between the flat spacetime and the boundary CFT which is the so-called ``flat holography". A natural way is to regard the Lorentz group in 4D as the global conformal group of a 2D CFT, which is known as "celestial holography". The most important element of celestial holography is the celestial amplitude \cite{Pasterski:2017kqt}, which tells us that we can expand the 4D flat amplitudes on a special basis, and then the amplitudes will behave as a CFT correlator in 2D. There are some good reviews about it \cite{Raclariu:2021zjz,Pasterski:2021rjz}.

In recent years, many new relations and new tools of celestial amplitudes have been developed. New bases are studied \cite{Chang:2022jut,Freidel:2022skz,Cotler:2023qwh}, new methods are discussed \cite{Hu:2022bpa,Tao:2023wls,Chang:2023ttm}, new concepts are proposed \cite{Melton:2023bjw,Jiang:2022hho,Strominger:2021lvk}, and so on. In addition, people also found some examples of the concrete structure of the boundary theory \cite{Costello:2022jpg,Costello:2023hmi,Stieberger:2022zyk,Stieberger:2023fju}. Meanwhile, the celestial versions of some special theories also caught our attention. In \cite{Banerjee:2023bni}, the authors studied celestial self-dual gravity theory, which is a $\omega_{1+\infty}$-invariant theory \cite{Banerjee:2023zip}, and found that the spectrum of the OPE may be bounded from below which is a non-trivial property. In this paper, we want to study the self-dual Yang-Mills theory (SDYM). Celestial SDYM theory is an $S$-invariant theory \cite{Banerjee:2023jne}, which has similar properties to self-dual gravity. For example, they both have the special property that they are one-loop exact \cite{Krasnov:2016emc}, which means that we can obtain the full spectrum when we consider the OPE limit of the celestial SDYM amplitudes\footnote{In fact, the same helicity one-loop amplitudes of the full YM are correctly captured by the SDYM. See \cite{Albayrak:2020saa} for some results of 1-loop celestial YM amplitudes.}.

In this paper, we derive a new formula to compute the celestial SDYM color-ordered amplitudes. This new formula can be expressed by two celestial Berends-Giele (BG) currents \cite{Tao:2023wls} connected to an effective propagator, which makes the leading OPE manifest since the OPE limit of the celestial BG currents is manifest. We also derive the OPE limit of the 5-pt celestial color-ordered SDYM amplitude, which can be connected to the OPE by the relation between the full amplitude and the color-ordered amplitude \cite{Fan:2019emx}. This paper will organized as follows. In section \ref{sec2}, we will review some concepts including celestial amplitudes, celestial BG currents, and the SDYM amplitudes. In section \ref{sec3}, we will derive the new formula for the celestial color-ordered SDYM amplitudes and show an example. In section \ref{sec4}, we will show how to obtain the leading OPE limit of the celestial color-ordered amplitudes only from the OPE limit of the celestial BG currents. In section \ref{sec5}, we will show higher orders of the OPE limit from direct calculations. There are still many interesting problems, we list them in the outlooks and leave them to future work.

\section{Preliminaries}\label{sec2}

\subsection{Celestial amplitudes and celestial BG currents}
In this subsection, we will review some basic concepts of celestial amplitudes and celestial BG currents. Celestial amplitudes can be obtained by expanding position space amplitudes $M(X_{i})$ with respect to the conformal primary wave functions $\phi_{\Delta,s}^{\pm}(z,\bar{z};X)$ (or shadow conformal primary wave functions $\tilde{\phi}^{\pm}_{\Delta,s}(z,\bar{z};X)$) instead of plane waves:
\ie\label{cdef}
\ma{A}^{\Delta_{i}}_{s_{i}}(z_{i},\bar{z}_{i})=(\prod_{j=1}^{n} \int d^4X_{j})( \prod_{j=1}^{k}\phi^{+}_{\Delta_{j},s_{j}}(z_{j},\bar{z}_{j};X_{j}))(\prod_{j=k+1}^{n} \tilde{\phi}^{-}_{\Delta_{j},s_{j}}(z_{j},\bar{z}_{j};X_{j}))A(X_{i}).
\fe
Here $\Delta$ and $s$ are the conformal dimension and the spin (or helicity) respectively, and $\pm$ labels whether the particle is outgoing or incoming. In this paper, we only focus on the Klein space with (2,2) signature where the incoming state and the outgoing state for the same particle can be connected by Lorentz transformation \cite{Pasterski:2017ylz}.

The coordinates $(z_{i},\bar{z}_{i})$ on the celestial sphere can be connected to massless on-shell momenta through
\ie
k^{\mu}=\epsilon\omega\hat{q}^{\mu}=\epsilon\omega(1+z\bar{z},z+\bar{z},-i(z-\bar{z}),1-z\bar{z})
\fe
and for massive on-shell momenta we have
\ie
k^{\mu}=\epsilon m\hat{p}^{\mu}=\epsilon\frac{m}{2y}(1+y^2+w\bar{w},w+\bar{w},-i(w-\bar{w}),1-y^2-w\bar{w})
\fe
Note that $\omega$ and $m$ are always positive and $\epsilon$ is $\pm1$ for outcoming and incoming particles respectively. In the Klein space, the primary wave functions for the massless particles can be regarded as the Mellin transformation of plane waves. We have the following formula which is more practical than equation \eqref{cdef}:
\ie
\ma{A}^{\Delta_{i}}_{s_{i}}(z_{i},\bar{z}_{i})=\int_{\omega_1,\cdots\omega_n\geqslant0}\prod_{i=1}^n\omega_i^{\Delta_i-1}A(q_{i})\delta^{(4)}(\sum_{i=1}^nq_i).
\fe

We can also derive the celestial amplitudes from celestial BG currents. The n-pt celestial BG current has the following formalism in the Klein space for massless particles (e.g. scalar off-shell)
\ie
\mathcal{J}(P)(\Delta_{i},p_{n+1})=\int \prod_{i=1}^{n} d\omega_{i}\omega_{i}^{\Delta_{i}-1}J(P)\delta^{(4)}(\sum_{i=1}^{n}q_{i}+p_{n+1})
\fe
where $J(P)$ is the BG current in the flat spacetime and can be derived recursively and $P$ denotes a set of on-shell momenta $P=k_{i_1},\cdots,k_{i_n}$. Note that here we can parameterize the off-shell massless momentum as a massive momentum. In the next subsection, we will show in a concrete theory how to obtain the BG currents recursively and finally write down the closed form.

\subsection{Amplitudes of SDYM}
SDYM is a truncation of the full YM theory that keeps only a subset of the Feynman diagrams of the latter. For some processes, for example, those involving mostly gluons of the same helicity, only the Feynman diagrams already present in SDYM can be seen to contribute in full YM calculations. This means that some of the processes can be computed in the simpler SDYM theory. In particular, the same helicity one-loop amplitudes of the full YM are correctly captured by the SD theory. One of the useful formulations of SDYM is given by Chalmers and Siegel \cite{Chalmers:1996rq}. In 2-component spinor notations, the action reads 
\be
S[\Psi, A] = \int \Psi^{a\, AB} (\partial_{A}{}^{A'} A^a_{A'B} + f^{abc} A^b_{A}{}^{A'} A^c_{A'B}).
\ee
The field $\Psi^{a\, AB}$ is a symmetric rank two spinor $\Psi^{a\, AB}= \Psi^{a\, (AB)}$ which takes values in the Lie algebra of some compact group $G$, and $a,b,c$ are the Lie algebra indices. The capital Latin letters (unprimed and primed) are the 2-component Lorentz spinor indices. The field $\Psi^{a\, AB}$ is an auxiliary field. If we take the variation of the action with respect to this field, it imposes the equation that the self-dual part of the field strength of the connection $A^a_{AA'}$ vanishes. This action can be linearised around any self-dual background. For simplicity, we expand it around the flat background. 
Consider the variation of the linearised action with respect to the auxiliary field. This gives the field equation $\partial_{A}{}^{A'} A^a_{A'B}=0$. The solutions describe one of the two gluon helicity, in our conventions positive. The corresponding polarisation vector is
\be
\epsilon_+^{MM'}(k)= \frac{q^M k^{M'}}{ \langle kq\rangle}.
\ee
Here $\langle \mu\nu\rangle = \mu^M \nu_M$ is the spinor contraction of two 2-component spinors $\mu^M, \nu^M$. The object $q^M$ is an auxiliary spinor changing which amounts to gauge-transformations. The object $k^{M'}$ is the momentum spinor arising as the square root of a null momentum $k^{MM'} = k^M k^{M'}$. 
\subsubsection{BG current for all positive helicity states}
The BG current can be defined to be the sum of all color-ordered Feynman diagrams, with positive helicity states of momenta $k_1, \cdots, k_n$ and the last leg kept off-shell. In our convention, the propagator is included in the off-shell leg of this current. Using recursion relation, it is not hard to show that the general n-point current takes the following form
\be\label{BG-vector}
J^{MM'}(k_1,\cdots, k_n) = J(k_1,\cdots, k_n) q^N (\sum_{i=1}^n k_{i\, N}{}^{M'}) q^M \equiv J(k) \langle q| \langle q| \sum_i k_i | ,
\ee
where
\be\label{YM-BG}
J(k_1,\cdots, k_n) = \frac{1}{\langle q1\rangle \langle 12\rangle \langle 23\rangle \cdots \langle (n-1) n\rangle \langle nq\rangle},
\ee
and we use the convention in which $k_1\equiv 1$ etc. 
\subsubsection{One-loop amplitudes}
The same helicity one-loop amplitudes at any multiplicity can be written as a sum over partitions of the set of external states into two groups.
Each group of states consists of a BG current, and two such currents are glued by the effective propagator. The amplitude is color-ordered, and so the states in one of the groups can be numbered
as those starting with the external particle $i$ and ending with particle $j$. This gives the following
amplitude \cite{Chattopadhyay:2021udc}
\\~\\
\begin{fmffile}{diagram}
\be 
\label{YM-general1}
\mathcal{A}=\sum_{part} \begin{gathered}
~~~\begin{fmfgraph*}(120,70)
     \fmfleft{i,i1,i2,i3,i4}
     \fmfright{o1,o4,o5,o6,o7}
     \fmf{vanilla}{i,v}
     \fmf{vanilla}{i4,v}
     \fmfblob{.15w}{v}
     \fmf{vanilla,label=$p_j$,label.distance=1cm,label.side=left}{v,v1}
     \fmfdot{v1}
     \fmf{vanilla,label=$p_{i-1}$,label.distance=1cm,label.side=right}{v1,o}
     \fmfblob{.15w}{o}
     \fmf{vanilla}{o,o1}
     \fmf{vanilla}{o,o7}
     \fmflabel{$i$}{i}
      \fmflabel{$.$}{i1}
     \fmflabel{$.$}{i2}
      \fmflabel{$.$}{i3}
      \fmflabel{$j$}{i4}
     \fmflabel{$i-1$}{o1}
      \fmflabel{$.$}{o4}
      \fmflabel{$.$}{o5}
      \fmflabel{$.$}{o6}
     \fmflabel{$j+1$}{o7}    
     \end{fmfgraph*}
     \end{gathered}
     \quad\quad
\ee 
\be 
=\sum_{part}J(i,\cdots,j)J(j+1,\cdots,i-1)\langle q|p_j\circ (k_i+\cdots+k_j)|q\rangle\langle q|(k_{j+1}+\cdots+k_{i-1})\circ p_{i-1}|q\rangle.\nonumber 
\ee 
Using 
\be 
k_i+\cdots+k_j=p_j-p_{i-1}=-(k_{j+1}+\cdots+k_{i-1}),
\ee 
we can write this compactly as 
\be 
\label{BG}
\mathcal{A}=\sum_{part}J(i,\cdots,j)J(j+1,\cdots,i-1)\langle q|p_j\circ p_{i-1}|q\rangle^2.
\ee 
It is also possible to write the above as a sum over cyclic permutations of the set $1,\cdots, n$.
\be\label{formula-cyclic}
2 {\cal A} = \sum_{i=1}^{n-1} J(1,\cdots, i) J(i+1,\cdots,n) \langle q| p_i \circ ( k_{i+1} + \cdots + k_n)| q\rangle^2 + {\rm cyclic},
\ee
where we need to add all cyclic permutations of the set $(1,\cdots, n)$.
\subsubsection{Color-dressed amplitude}
The formula for the SDYM amplitude we wrote above is valid for a color-ordered amplitude. We can obtain the full color-dressed amplitude using the following formula \cite{Monteiro:2022nqt}. 
\ie\label{famp}
\mathcal{A}_{\text{SDYM}}=\sum_{\sigma\in S_{n-1}/\mathcal{R}}c^{a_{\sigma(1)}a_{\sigma(2)}a_{\sigma(3)}\cdots a_{\sigma(n)}}\mathcal{A}(\sigma(1)\sigma(2)\sigma(3)\cdots\sigma(n)).
\fe 
where the sum is over non-cyclic permutations, modulo reflection of the list $\sigma$. The color factor is given by
\ie
c^{a_1a_2\cdots a_n}=f^{b_1a_1b_2}f^{b_2a_2b_3}\cdots f^{b_na_nb_1},
\fe
where $f^{abc}$ is the structure constant of the gauge group.

\section{A new formula for celestial SDYM}\label{sec3}
In this section, we will give a new formula for celestial SDYM amplitudes based on the formula for SDYM amplitudes. Then we will show an example. The application of this new formula will be shown in the next section.
\subsection{A new formula}
Recall that the 1-loop amplitude of SDYM can be written as 
\ie
A_n=&\sum_{part}J(i,\cdots,j)J(j+1,\cdots,i-1)\langle q|p_j\circ p_{i-1}|q\rangle^2\delta(\sum k_i)=\\
&\sum_{part}\int dk_Ldk_R\ J(i,\cdots,j)\delta(\sum_{k=i}^jk_k+k_L)J(j+1,\cdots,i-1)\delta(\sum_{k=j+1}^{i-1}k_k+k_R)\langle q|p_j\circ p_{i-1}|q\rangle^2\delta(k_L+k_R)
\fe
where $k_L$($k_R$) is the momentum of the left(right) off-shell leg. An alternative formula is
\ie\label{sdyma}
2A_n=\sum_{i=1}^{n-1}J(1,2,\cdots,i)J(i+1,\cdots,n)\langle q|p_i\circ k_R|q\rangle^2\delta(\sum k_i)+\text{cyclic}.
\fe
For later convenience, we choose the following parameterization for the region momenta:
\ie
p_i=k_n+\sum_{k=1}^{i}k_k+x
\fe
Since the SDYM is independent of $x$, we can drop all terms containing $x$. Then the effective propagator is given by
\ie
\langle q|p_i\circ k_R|q\rangle^2\sim\langle q|k_n+\sum_{k=1}^{i}k_k\circ k_R|q\rangle^2=\langle q|\sum_{k=n}^{i}k_k\circ k_R|q\rangle^2
\fe
where we assume that the number after $n$ is 1. From the definition of the n-pt celestial BG current
\ie
\mathcal{J}(P)(\Delta_{i},k_{n+1})=\int \prod_{i=1}^{n} d\omega_{i}\omega_{i}^{\Delta_{i}-1}J(P)\delta^{(4)}(\sum_{i=1}^{n}k_n+k_{n+1}),
\fe
we can translate the SDYM amplitude formula \eqref{sdyma} to a celestial SDYM formula\footnote{Note that now the $k_R$ in the effective propagator should be expressed as massive spinors. The conventions and the derivation of the effective propagator can be found in appendix \ref{app1}.}:
\ie
2\sum_{i=1}^{n-1}\int dk_Ldk_R\ \ma{J}(i+1,\cdots,n)(k_R)\frac{m_R^2}{y_R^2}\sum_{k,l=n}^{i}\epsilon_k\epsilon_l\ma{J}^{k,l}(1,\cdots,i)(k_L)\bar{z}_{kR}\bar{z}_{lR}\delta(k_L+k_R)+\text{cyclic},
\fe
where the notation $\ma{J}^{k,l}$ means that we take $\Delta_k\rightarrow\Delta_k+1$ and $\Delta_l\rightarrow\Delta_l+1$ in $\ma{J}$. Note that the massive momentum $k_R$ can be parameterized as
\ie
k_R=\epsilon_R\frac{m_R}{2y_R}(1+y_R^2+z_R\bar{z}_R,z_R+\bar{z}_R,-i(z_R-\bar{z}_R),1-y_R^2-z_R\bar{z}_R).
\fe
After integrating $k_L$ out and using the above parameterization for $k_R$, we will obtain the final formalism for celestial SDYM amplitudes:
\ie\label{ca}
\ma{A}_n=64\sum_{i=1}^{n-1}\int da_Rdy_Rd^2z_R\ a_R^5y_R\ma{J}(i+1,\cdots,n)(k_R)\sum_{k,l=n}^{i}\epsilon_k\epsilon_l\ma{J}^{k,l}(1,\cdots,i)(-k_R)\bar{z}_{kR}\bar{z}_{lR}+\text{cyclic}
\fe
where $a_R=m_R/2y_R$. 

Now we will give some celestial BG currents for SDYM as a warm-up. Using the identities
\ie
\langle ij\rangle=-2\epsilon_i\epsilon_j\sqrt{\omega_i\omega_j}z_{ij},[ij]=2\sqrt{\omega_i\omega_j}\bar{z}_{ij}
\fe
and the special choice of the reference spinor $|q\rangle=(0,1)$ and $\langle q|=(-1,0)$, we have the 1-pt celestial BG current for $k_1$ incoming and $k_R$ outgoing:
\ie
\ma{J}(1)(k_R)=-\frac{1}{2}\int d\omega_1 \omega_1^{\Delta_1-2}\delta(k_1+k_R)=-\frac{1}{16}\frac{a_R^{\Delta_1-5}}{y_R}\delta(y_R)\delta^{(2)}(z_R-z_1).
\fe
We can also obtain the 2-pt celestial BG current for $k_1$, $k_2$ incoming and $k_R$ outgoing:
\ie
\ma{J}(12)(k_R)=\int d\omega_1d\omega_2\omega_1^{\Delta_1-2}\omega_2^{\Delta_2-2}\frac{1}{4z_{12}}\delta^{(4)}(k_1+k_2+k_R).
\fe
After substituting
\ie
&\delta^{(4)}(k_1+k_2+k_R)=\frac{2y_R^4}{m_R^3(y_R^2+|z_{R2}|^2)}\delta(\omega_2-\frac{m_R^2}{4\omega_1|z_{12}|^2})\delta(y_R-\frac{2m_R\omega_1|z_{12}|^2}{m_R^2+4\omega_1^2|z_{12}|^2})\\
&\times\delta^{(2)}(z_R-\frac{m_R^2z_2+4\omega_1^2z_1|z_{12}|^2}{m_R^2+4\omega_1^2|z_{12}|^2})=(\frac{2m_R\omega_1|z_{12}|^2}{m_R^2+4\omega_1^2|z_{12}|^2})^3\frac{1}{m_R^2\omega_1|z_{12}|^2}\delta^{(4)},
\fe
we will get a more practical result:
\ie
\ma{J}(12)(k_R)=&-2^{7-2\Delta_2}\int d\omega_1\omega_1^{\Delta_1-\Delta_2+2}\frac{1}{z_{12}}\frac{m_R^{2\Delta_2-3}|z_{12}|^{8-2\Delta_2}}{(m_R^2+4\omega_1^2|z_{12}|^2)^3}\\
&\times\delta(y_R-\frac{2m_R\omega_1|z_{12}|^2}{m_R^2+4\omega_1^2|z_{12}|^2})\delta^{(2)}(z_R-\frac{m_R^2z_2+4\omega_1^2z_1|z_{12}|^2}{m_R^2+4\omega_1^2|z_{12}|^2}).
\fe
Furthermore, the most general celestial BG current is given by
\ie
\ma{J}(12\cdots n)(k_R)=\frac{1}{(-2)^nz_{12}z_{23}\cdots z_{n-1,n}}\int_{\omega_1,\cdots,\omega_n}\prod_{i=1}^{n}\omega_i^{\Delta_i-2}\delta^{(4)}(\sum_{i=1}^nk_i+k_R).
\fe

\subsection{Example: 4-pt celestial amplitude of SDYM}
For loop amplitudes, the 3-pt case will give a vanishing amplitude. The first non-trivial case is the 4-pt case. In this subsection, we will show how to derive the 4-pt celestial amplitude in SDYM using our new formula.

Now we consider the $2\rightarrow2$ celestial amplitude. We set $k_1$, $k_2$ incoming and $k_3$, $k_4$ outgoing. In order to simplify the calculation, we will choose the reference spinor $q\propto k_3$. This celestial amplitude can be written as
\ie
\ma{A}_{2\rightarrow2}=&64\sum_{i=1}^{3}\sum_{k,l=4}^{i}\int da_Rdy_Rd^2z_R\ a_R^5y_R\ma{J}(i+1,\cdots,n)(k_R)\ma{J}^{k,l}(1,\cdots,i)(-k_R)\Pi_{k,l}+\text{cyclic}\\
=&64\sum_{k,l=4}^{1}\int da_Rdy_Rd^2z_R\ a_R^5y_R\ma{J}(2,3,4)(k_R)\epsilon_k\epsilon_l\ma{J}^{k,l}(1)(-k_R)\Pi_{k,l}\\
&+64\sum_{k,l=4}^{2}\int da_Rdy_Rd^2z_R\ a_R^5y_R\ma{J}(3,4)(k_R)\epsilon_k\epsilon_l\ma{J}^{k,l}(1,2)(-k_R)\Pi_{k,l}\\
&+64\sum_{k,l=4}^{3}\int da_Rdy_Rd^2z_R\ a_R^5y_R\ma{J}(4)(k_R)\epsilon_k\epsilon_l\ma{J}^{k,l}(1,2,3)(-k_R)\Pi_{k,l}+\text{cyclic}.
\fe
After using the momentum conservation, we have
\ie
&\ma{A}_{2\rightarrow2}=128\int da_Rdy_Rd^2z_R\ a_R^5y_R\ma{J}^{4,4}(2,3,4)(k_R)\ma{J}(1)(-k_R)\Pi_{4,4}\\
&+128\int da_Rdy_Rd^2z_R\ a_R^5y_R\ma{J}^{3,3}(3,4,1)(k_R)\ma{J}(2)(-k_R)\Pi_{3,3}\\
&+64\int da_Rdy_Rd^2z_R\ a_R^5y_R\ma{J}^{3,3}(3,4)(k_R)\ma{J}(1,2)(-k_R)\Pi_{3,3}\\
&+64\int da_Rdy_Rd^2z_R\ a_R^5y_R\ma{J}(1,2)(k_R)\ma{J}^{4,4}(3,4)(-k_R)\Pi_{4,4}\\
=&128\int da_Rdy_Rd^2z_R\ a_R^5y_R\ma{J}^{4,4}(2,3,4)(k_R)\ma{J}(1)(-k_R)\Pi_{4,4}\\
&+128\int da_Rdy_Rd^2z_R\ a_R^5y_R\ma{J}^{3,3}(3,4,1)(k_R)\ma{J}(2)(-k_R)\Pi_{3,3}\\
&+128\int da_Rdy_Rd^2z_R\ a_R^5y_R\ma{J}^{3,3}(3,4)(k_R)\ma{J}(1,2)(-k_R)\Pi_{3,3}
\fe
where we have used
\ie
\langle q3\rangle[3k_R^a]\langle k_{Ra}q\rangle+\langle q4\rangle[4k_R^a]\langle k_{Ra}q\rangle=0
\fe
from
\ie
|3\rangle[3|+|4\rangle[4|=|k^a_R\rangle[k_{Ra}|
\fe
for the celestial BG current $\ma{J}(3,4)(-k_R)$. Since $\Pi_{3,3}=0$ is a zero of order 2, the 4-pt celestial amplitude is given by
\ie
\ma{A}_{2\rightarrow2}=&128\int da_Rdy_Rd^2z_R\ a_R^5y_R\ma{J}^{4,4}(2,3,4)(k_R)\ma{J}(1)(-k_R)\Pi_{4,4}.
\fe
Note that this time we have the reference spinor $q=k_3$. Then we have
\ie
\ma{A}_{2\rightarrow2}=&-\frac{\bar{z}^2_{14}}{z_{23}^2}\int_{\omega_1,\omega_2,\omega_3,\omega_4}\omega_1^{\Delta_1}\omega_2^{\Delta_2-2}\omega_3^{\Delta_3-2}\omega_4^{\Delta_4}\delta^{(4)}(\sum_{i=1}^4k_i).
\fe
From
\ie
&\delta^{(4)}(-\omega_1\hat{q}_1-\omega_2\hat{q}_2+\omega_3\hat{q}_3+\omega_4\hat{q}_4)=\\
&\frac{1}{4\omega_4}\delta(\omega_1-\omega_4\frac{z_{24}\bar{z}_{34}}{z_{12}\bar{z}_{13}})\delta(\omega_2+\omega_4\frac{z_{14}\bar{z}_{34}}{z_{12}\bar{z}_{23}})\delta(\omega_3+\omega_4\frac{z_{24}\bar{z}_{14}}{z_{23}\bar{z}_{13}})\delta(z_{12}z_{34}\bar{z}_{13}\bar{z}_{24}-z_{13}z_{24}\bar{z}_{12}\bar{z}_{34}),
\fe
we have
\ie
\ma{A}_{2\rightarrow2}=&-\frac{\bar{z}^2_{14}}{z_{23}^2}\int_{\omega_1,\omega_2,\omega_3,\omega_4}\omega_1^{\Delta_1}\omega_2^{\Delta_2-2}\omega_3^{\Delta_3-2}\omega_4^{\Delta_4}\delta^{(4)}(\sum_{i=1}^4k_i)\\
=&-\frac{\bar{z}^2_{14}}{2z_{23}^2}\pi\delta(\sum_{i=1}^4\lambda_i)(\frac{z_{24}\bar{z}_{34}}{z_{12}\bar{z}_{13}})^{\Delta_1}(\frac{z_{41}\bar{z}_{34}}{z_{12}\bar{z}_{23}})^{\Delta_2-2}(\frac{z_{24}\bar{z}_{41}}{z_{23}\bar{z}_{13}})^{\Delta_3-2}\delta(z_{12}z_{34}\bar{z}_{13}\bar{z}_{24}-z_{13}z_{24}\bar{z}_{12}\bar{z}_{34})\\
=&\frac{1}{2}\pi\delta(\sum_{i=1}^4\lambda_i)(\frac{z_{24}\bar{z}_{34}}{z_{12}\bar{z}_{13}})^{\Delta_1-1}(\frac{z_{41}\bar{z}_{34}}{z_{12}\bar{z}_{23}})^{\Delta_2}(\frac{z_{24}\bar{z}_{41}}{z_{23}\bar{z}_{13}})^{\Delta_3-1}\frac{\bar{z}_{12}\bar{z}_{23}}{z_{14}z_{34}}\delta(z_{12}z_{34}\bar{z}_{13}\bar{z}_{24}-z_{13}z_{24}\bar{z}_{12}\bar{z}_{34})
\fe
using the condition that $\frac{z_{12}\bar{z}_{23}}{z_{41}\bar{z}_{34}}$ is real. Note that we focus on the principle series $\Delta_i=1+i\lambda_i$ and we have used the following identity:
\ie
\int_0^{\infty} d\omega \omega^{i\lambda-1}=2\pi\delta(\lambda).
\fe

\section{Leading OPE limit from the new formula}\label{sec4}
In this section, we will show the leading OPE of the celestial BG currents and then show how to obtain the OPE of the celestial SDYM amplitudes from it.
\subsection{OPE limit of the celestial BG currents}
There is a standard analysis of the relation between the collinear limit and the OPE limit in \cite{Pate:2019lpp}. This analysis is easy to be generalized to our case. In SDYM, the collinear limit for adjacent two legs of a certain amplitude can be expressed as the collinear limit of BG currents: 
\ie\label{col}
J(1,2,\cdots,j,j+1,\cdots,n)\rightarrow\frac{1}{\sqrt{t(1-t)}\langle j,j+1\rangle}J(1,2,\cdots,P,\cdots,n)
\fe
where $t$ is a parameter that describes how the two momentum $k_j$ and $k_{j+1}$ become collinear, i.e. $k_j\rightarrow tk_P$ and $k_{j+1}\rightarrow (1-t)k_P$. Assuming they are all outgoing, we can obtain the leading OPE of celestial amplitudes by considering the following integral which comes from the Mellin transformation of the BG current:
\ie
-\int d\omega_j\int d\omega_{j+1}\omega_j^{\Delta_j-3/2}\omega_{j+1}^{\Delta_{j+1}-3/2}\frac{1}{2\sqrt{t(1-t)}z_{j,j+1}}\cdots,
\fe
which will give\footnote{In \cite{Tao:2023wls}, the author does not use the collinear parameterization and therefore obtains an unusual result.}
\ie
&-\int d\omega_j\int d\omega_{j+1}\omega_j^{\Delta_j-3/2}\omega_{j+1}^{\Delta_{j+1}-3/2}\frac{1}{2\sqrt{t(1-t)}z_{j,j+1}}\cdots\\
&=-\frac{1}{2z_{j,j+1}}\int_0^1 dt\int d\omega_P \omega_P^{\Delta_j+\Delta_{j+1}-2}t^{\Delta_j-2}(1-t)^{\Delta_{j+1}-2}\cdots\\
&=-\frac{1}{2z_{j,j+1}}B(\Delta_j-1,\Delta_{j+1}-1)\int d\omega_P \omega_P^{\Delta_j+\Delta_{j+1}-2}\cdots.
\fe
Note that the higher-order terms of the OPE limit will not appear in this way since it requires that the delta function is invariant under the collinear limit which means that we have forced higher-order terms to vanish. We can conclude that the leading OPE limit of the celestial BG currents is (we use the $2\rightarrow3$ celestial amplitude as an example)
\ie\label{lope}
&\ma{A}_{2\rightarrow3}\sim-\frac{B(\Delta_4-1,\Delta_5-1)}{2z_{45}}\ma{A}_{2\rightarrow2}^{\Delta_4\rightarrow\Delta_4+\Delta_5-1}+\cdots.
\fe
However, there are two BG currents in the formula \eqref{sdyma} and one may be concerned about the case that these two adjacent legs belong to different BG currents. This case will not appear if we choose a suitable parameterization for the region momenta. If we want to consider the collinear limit of $j$ and $j+1$, we only need to set $p_j=x$ in the parameterization. Under this choice, we only need to consider the case that two adjacent legs belong to the same BG current, and the analysis above is valid. One may also doubt whether the leading OPE limit \eqref{lope} is correct in the SDYM case since the relation between the BG currents and the SDYM amplitudes is not as direct as the usual YM theory. In the next subsection, we will compute the $2\rightarrow3$ celestial amplitude case to verify the discussion in this subsection.

\subsection{Leading OPE limit for 5-pt celestial color-ordered SDYM amplitudes}
To verify our discussion, calculating an example will be helpful. We will consider the OPE limit of the 5-pt celestial amplitude without giving explicit formalism of this celestial amplitude.

From our formula, the 5-pt celestial amplitude is given by
\ie
\ma{A}_{2\rightarrow3}=&128\int da_Rdy_Rd^2z_1\ a_R^5y_R\ma{J}^{5,5}(2,3,4,5)(k_R)\ma{J}(1)(-k_R)\Pi_{5,5}\\
&+128\int da_Rdy_Rd^2z_R\ a_R^5y_R\ma{J}(1,2)(k_R)\ma{J}^{5,5}(3,4,5)(-k_R)\Pi_{5,5}\\
&+128\int da_Rdy_Rd^2z_R\ a_R^5y_R\ma{J}(1,2,3)(k_R)\ma{J}^{5,5}(4,5)(-k_R)\Pi_{5,5}\\
&+128\int da_Rdy_Rd^2z_R\ a_R^5y_R\ma{J}^{4,4}(4,5,1)(k_R)\ma{J}(2,3)(-k_R)\Pi_{4,4}\\
&+128\int da_Rdy_Rd^2z_R\ a_R^5y_R\ma{J}^{4,4}(4,5,1,2)(k_R)\ma{J}(3)(-k_R)\Pi_{4,4}\\
&+128\int da_Rdy_Rd^2z_R\ a_R^5y_R\ma{J}^{4,4}(3,4,5,1)(k_R)\ma{J}(2)(-k_R)\Pi_{4,4}.
\fe
We take the collinear limit \eqref{col} for the legs $4$ and $5$, and then replace the label $P$ with $4$ to make the leading OPE more transparent:
\ie
\ma{A}_{2\rightarrow3}\rightarrow&-64\frac{B(\Delta_4-1,\Delta_5+1)}{z_{45}}\int da_Rdy_Rd^2z_1\ a_R^5y_R\ma{J}(2,3,4^{\Delta_4+\Delta_5+1})(k_R)\ma{J}(1)(-k_R)\Pi_{4,4}\\
&-64\frac{B(\Delta_4+1,\Delta_5-1)}{z_{45}}\int da_Rdy_Rd^2z_R\ a_R^5y_R\ma{J}(4^{\Delta_4+\Delta_5+1},1)(k_R)\ma{J}(2,3)(-k_R)\Pi_{4,4}\\
&-64\frac{B(\Delta_4+1,\Delta_5-1)}{z_{45}}\int da_Rdy_Rd^2z_R\ a_R^5y_R\ma{J}(4^{\Delta_4+\Delta_5+1},1,2)(k_R)\ma{J}(3)(-k_R)\Pi_{4,4}\\
&-64\frac{B(\Delta_4+1,\Delta_5-1)}{z_{45}}\int da_Rdy_Rd^2z_R\ a_R^5y_R\ma{J}(3,4^{\Delta_4+\Delta_5+1},1)(k_R)\ma{J}(2)(-k_R)\Pi_{4,4}\\
&-128\frac{B(\Delta_4,\Delta_5-1)}{z_{45}}\int da_Rdy_Rd^2z_R\ a_R^5y_R\ma{J}^{3}(3,4^{\Delta_4+\Delta_5},1)(k_R)\ma{J}(2)(-k_R)\Pi_{3,4}\\
=&-64\frac{B(\Delta_4-1,\Delta_5+1)}{z_{45}}\int da_Rdy_Rd^2z_1\ a_R^5y_R\ma{J}(2,3,4^{\Delta_4+\Delta_5+1})(k_R)\ma{J}(1)(-k_R)\Pi_{4,4}\\
&-64\frac{B(\Delta_4+1,\Delta_5-1)}{z_{45}}\int da_Rdy_Rd^2z_R\ a_R^5y_R\ma{J}(4^{\Delta_4+\Delta_5+1},1)(k_R)\ma{J}(2,3)(-k_R)\Pi_{4,4}\\
&-64\frac{B(\Delta_4+1,\Delta_5-1)}{z_{45}}\int da_Rdy_Rd^2z_R\ a_R^5y_R\ma{J}(4^{\Delta_4+\Delta_5+1},1,2)(k_R)\ma{J}(3)(-k_R)\Pi_{4,4}\\
&-64\frac{B(\Delta_4+1,\Delta_5-1)}{z_{45}}\int da_Rdy_Rd^2z_R\ a_R^5y_R\ma{J}^{1,1}(3,4^{\Delta_4+\Delta_5-1},1)(k_R)\ma{J}(2)(-k_R)\Pi_{1,1}\\
&+128\frac{B(\Delta_4,\Delta_5-1)}{z_{45}}\int da_Rdy_Rd^2z_R\ a_R^5y_R\ma{J}^{3}(3,4^{\Delta_4+\Delta_5},1)(k_R)\ma{J}(2)(-k_R)\Pi_{1,3}\\
&-128\frac{B(\Delta_4+1,\Delta_5-1)}{z_{45}}\int da_Rdy_Rd^2z_R\ a_R^5y_R\ma{J}^{1,3}(3,4^{\Delta_4+\Delta_5-1},1)(k_R)\ma{J}(2)(-k_R)\Pi_{1,3}.
\fe
Note that if we choose $p_4=x$, the 4-pt celestial amplitude $\ma{A}_{2\rightarrow2}$ can also be written as
\ie
\ma{A}_{2\rightarrow2}=&128\int da_Rdy_Rd^2z_R\ a_R^5y_R\ma{J}^{1,1}(3,4,1)(k_R)\ma{J}(2)(-k_R)\Pi_{1,1}\\
&+128\int da_Rdy_Rd^2z_R\ a_R^5y_R\ma{J}^{1,1}(4,1)(k_R)\ma{J}(2,3)(-k_R)\Pi_{1,1}\\
&+128\int da_Rdy_Rd^2z_R\ a_R^5y_R\ma{J}^{4,4}(4,1,2)(k_R)\ma{J}(3)(-k_R)\Pi_{4,4}.
\fe
In addition, from the direct calculation, we have
\ie
\int da_Rdy_Rd^2z_R\ a_R^5y_R\ma{J}^{1,3}(3,4^{\Delta_4+\Delta_5-1},1)(k_R)\ma{J}(2)(-k_R)\Pi_{1,3}=-\ma{A}^{\Delta_4\rightarrow\Delta_4+\Delta_5-1}_{2\rightarrow2}.
\fe
Then the leading OPE limit of 5-pt celestial color-ordered amplitude is
\ie
\ma{A}_{2\rightarrow3}\rightarrow-\frac{B(\Delta_4-1,\Delta_5-1)}{2z_{45}}\ma{A}^{\Delta_4\rightarrow\Delta_4+\Delta_5-1}_{2\rightarrow2},
\fe
which gives the leading OPE limit \eqref{lope}. Using \eqref{famp} we can also get the leading OPE. In this case, we only need to consider the color-ordered amplitudes with the legs 4 and 5 adjacent. The result is
\ie
\ma{O}^a_{\Delta_4}(z_4)\ma{O}^b_{\Delta_5}(z_5)\sim-\frac{f^{abc}B(\Delta_4-1,\Delta_5-1)}{2z_{45}}\ma{O}^c_{\Delta_4+\Delta_5-1}(z_5),
\fe
where we can ignore the difference between $z_4$ and $z_5$ at the leading order. This matches with the usual result for the YM theory up to an overall normalization factor.

\section{Higher-order OPE terms from the direct calculation}\label{sec5}
In this section, we want to explore the higher-order terms in the OPE limit to make our study on the celestial SDYM more complete. We will calculate the 5-pt celestial amplitudes directly and figure out such terms. We will also double-check the result in section \ref{sec3} and section \ref{sec4}. For later convenience, we will replace the label $4$ with $5$ in the 4-pt celestial amplitude $\mathcal{A}_{2\rightarrow2}$ in this section. Let us write down the $2\rightarrow2$ celestial amplitude first:
\ie
\ma{A}_4(1,2,3,5)&=\frac{s_{12}s_{23}}{\langle12\rangle\langle23\rangle\langle35\rangle\langle15\rangle}\\
\ma{A}_{2\rightarrow2}&=\int_{\omega_1,\omega_2,\omega_3,\omega_5}\omega_1^{\Delta_1-1}\omega_2^{\Delta_2}\omega_3^{\Delta_3-1}\omega_5^{\Delta_5-2}\frac{\bar{z}_{12}\bar{z}_{23}}{z_{35}z_{15}}\delta^{(4)}(-\omega_1\hat{q}_1-\omega_2\hat{q}_2+\omega_3\hat{q}_3+\omega_5\hat{q}_5).
\fe
From
\ie
&\delta^{(4)}(-\omega_1\hat{q}_1-\omega_2\hat{q}_2+\omega_3\hat{q}_3+\omega_5\hat{q}_5)=\\
&\frac{1}{4\omega_5}\delta(\omega_1-\omega_5\frac{z_{25}\bar{z}_{35}}{z_{12}\bar{z}_{13}})\delta(\omega_2+\omega_5\frac{z_{15}\bar{z}_{35}}{z_{12}\bar{z}_{23}})\delta(\omega_3+\omega_5\frac{z_{25}\bar{z}_{15}}{z_{23}\bar{z}_{13}})\delta(z_{12}z_{34}\bar{z}_{13}\bar{z}_{25}-z_{13}z_{25}\bar{z}_{12}\bar{z}_{35}),
\fe
we have
\ie
\ma{A}_{2\rightarrow2}=&\frac{1}{4}\int_{\omega_5}(\omega_5\frac{z_{25}\bar{z}_{35}}{z_{12}\bar{z}_{13}})^{\Delta_1-1}(\omega_5\frac{z_{51}\bar{z}_{35}}{z_{12}\bar{z}_{23}})^{\Delta_2}(\omega_5\frac{z_{25}\bar{z}_{51}}{z_{23}\bar{z}_{13}})^{\Delta_3-1}\omega_5^{\Delta_5-3}\frac{\bar{z}_{12}\bar{z}_{23}}{z_{35}z_{15}}\delta(z_{12}z_{35}\bar{z}_{13}\bar{z}_{25}-z_{13}z_{25}\bar{z}_{12}\bar{z}_{35})\\
=&\frac{1}{2}\pi\delta(\sum_i\lambda_i)(\frac{z_{25}\bar{z}_{35}}{z_{12}\bar{z}_{13}})^{\Delta_1-1}(\frac{z_{51}\bar{z}_{35}}{z_{12}\bar{z}_{23}})^{\Delta_2}(\frac{z_{25}\bar{z}_{51}}{z_{23}\bar{z}_{13}})^{\Delta_3-1}\frac{\bar{z}_{12}\bar{z}_{23}}{z_{35}z_{15}}\delta(z_{12}z_{35}\bar{z}_{13}\bar{z}_{25}-z_{13}z_{25}\bar{z}_{12}\bar{z}_{35})
\fe
where $\Delta_i=1+i\lambda_i$. This result matches the result from our new formula. Then we write down the $2\rightarrow 3$ celestial amplitude. From
\ie
A_5=\frac{s_{12}s_{23}+s_{45}s_{51}+s_{25}s_{45}+s_{25}s_{14}+\langle24\rangle[14]\langle15\rangle[52]}{\langle12\rangle\langle23\rangle\langle34\rangle\langle45\rangle\langle51\rangle},
\fe
we have
\ie
&\ma{A}_{2\rightarrow3}=\frac{1}{2}\int_{\omega_{1,2,3,4,5}}\omega_1^{\Delta_1-2}\omega_2^{\Delta_2-2}\omega_3^{\Delta_3-2}\omega_4^{\Delta_4-2}\omega_5^{\Delta_5-2}\delta^{(4)}\times\\
&\frac{\omega_1\omega_2^2\omega_3|z_{12}|^2|z_{23}|^2+\omega_4\omega_5^2\omega_1|z_{45}|^2|z_{15}|^2+\omega_2\omega_5^2\omega_4|z_{25}|^2|z_{45}|^2-\omega_1\omega_2\omega_4\omega_5|z_{25}|^2|z_{14}|^2+\omega_1\omega_2\omega_4\omega_5z_{24}\bar{z}_{14}z_{15}\bar{z}_{25}}{z_{12}z_{23}z_{34}z_{45}z_{51}}
\fe
Since we are only interested in the collinear limit, we choose the following parameterization
\ie
\omega_4=t\omega_P,\omega_5=(1-t)\omega_P
\fe
and then the delta function is \cite{Banerjee:2023bni}
\ie
&\delta^{(4)}(\sum_{i=1}^5\epsilon_i\omega_i\hat{q}_i)=\frac{1}{4\omega_P}\delta(\omega_1-\omega_1^*)\delta(\omega_2-\omega_2^*)\delta(\omega_3-\omega_3^*)\\
&\times\delta\bigg(x-\bar{x}-tz_{45}(\frac{x}{z_{35}}-\frac{\bar{x}}{z_{25}})-t\bar{z}_{45}(\frac{x}{\bar{z}_{25}}-\frac{\bar{x}}{\bar{z}_{35}})+tz_{45}\bar{z}_{45}(\frac{x}{z_{35}\bar{z}_{25}}-\frac{\bar{x}}{z_{25}\bar{z}_{35}})\bigg)
\fe
where
\ie
x&=z_{12}z_{35}\bar{z}_{13}\bar{z}_{25},~~
\bar{x}=z_{13}z_{25}\bar{z}_{12}\bar{z}_{35}\\
\omega_i^*&=\epsilon_i\omega_P(\sigma_{i,1}+tz_{45}\sigma_{i,2}+t\bar{z}_{45}\sigma_{i,3}+tz_{45}\bar{z}_{45}\sigma_{i,4})\\
\sigma_{1,1}&=-\frac{z_{25}\bar{z}_{35}}{z_{12}\bar{z}_{13}},~~
\sigma_{2,1}=\frac{z_{15}\bar{z}_{35}}{z_{12}\bar{z}_{23}}\\
\sigma_{3,1}&=-\frac{z_{25}\bar{z}_{15}}{z_{23}\bar{z}_{13}},~~
\sigma_{i,2}=\frac{\partial\sigma_{i,1}}{\partial z_5}\\
\sigma_{i,3}&=\frac{\partial\sigma_{i,1}}{\partial \bar{z}_5},~~
\sigma_{i,4}=\frac{\partial^2\sigma_{i,1}}{\partial z_5\partial \bar{z}_5}.
\fe
Thus we can perform the integrals over $\omega_1,\omega_2,\omega_3$ to obtain 
\ie
\label{fivepoint}
&\ma{A}_{2\rightarrow3}=\frac{1}{8}\int_{\omega_{4,5}}\prod_i^3(\omega^*_i)^{\Delta_i-2}\omega_4^{\Delta_4-2}\omega_5^{\Delta_5-2}\times\\
&\frac{\omega^*_1\omega^{*2}_2\omega^*_3|z_{12}|^2|z_{23}|^2+\omega_4\omega_5^2\omega^*_1|z_{45}|^2|z_{15}|^2+\omega^*_2\omega_5^2\omega_4|z_{25}|^2|z_{45}|^2-\omega^*_1\omega^*_2\omega_4\omega_5|z_{25}|^2|z_{14}|^2+\omega^*_1\omega^*_2\omega_4\omega_5z_{24}\bar{z}_{14}z_{15}\bar{z}_{25}}{\omega_Pz_{12}z_{23}z_{34}z_{45}z_{51}}\\&\times\delta\bigg(x-\bar{x}-tz_{45}(\frac{x}{z_{35}}-\frac{\bar{x}}{z_{25}})-t\bar{z}_{45}(\frac{x}{\bar{z}_{25}}-\frac{\bar{x}}{\bar{z}_{35}})+tz_{45}\bar{z}_{45}(\frac{x}{z_{35}\bar{z}_{25}}-\frac{\bar{x}}{z_{25}\bar{z}_{35}})\bigg).
\fe
Now we have collected all the elements and next we will find the OPE limit of legs $4$ and $5$ of the $2\rightarrow3$ celestial amplitude.

\subsection{Leading order}
For the leading order, we have
\ie
\ma{A}_{2\rightarrow3}|_{\mathcal{O}(\frac{1}{z_{45}})}&=\frac{\pi}{4z_{45}}\delta(\sum_{i=1}^5\lambda_i)B(\Delta_4-1,\Delta_5-1)
\Big(\frac{z_{25}\bar{z}_{35}}{z_{12}\bar{z}_{13}}\Big)^{\Delta_1-1}\Big(\frac{z_{51}\bar{z}_{35}}{z_{12}\bar{z}_{23}}\Big)^{\Delta_2}\Big(\frac{z_{25}\bar{z}_{51}}{z_{23}\bar{z}_{13}}\Big)^{\Delta_3-1}\times\Big(\frac{\bar{z}_{12}\bar{z}_{23}}{z_{35}z_{51}}\Big)\delta(x-\bar{x})\\
&=-\frac{1}{2z_{45}}\mathcal{A}_{2\rightarrow2}^{\Delta_5\rightarrow\Delta_4+\Delta_5-1},
\fe
which matches with the result before. 
 
\subsection{Terms at the order $\ma{O}(\frac{\bar{z}_{45}}{z_{45}})$}
We first write down the derivative to $\bar{z}_4$ of $\ma{A}_{2\rightarrow2}$.
\ie
\label{derivative}
\bar{\partial}_5\ma{A}_{2\rightarrow2}=&-\frac{1}{2}\pi\delta(\sum_i\lambda_i)(\Delta_1-1)(\frac{z_{25}}{z_{12}\bar{z}_{13}})(\frac{z_{25}\bar{z}_{35}}{z_{12}\bar{z}_{13}})^{\Delta_1-2}(\frac{z_{51}\bar{z}_{35}}{z_{12}\bar{z}_{23}})^{\Delta_2}(\frac{z_{25}\bar{z}_{51}}{z_{23}\bar{z}_{13}})^{\Delta_3-1}\frac{\bar{z}_{12}\bar{z}_{23}}{z_{35}z_{15}}\delta(x-\bar{x})\\
&-\frac{1}{2}\pi\delta(\sum_i\lambda_i)\Delta_2(\frac{z_{25}\bar{z}_{35}}{z_{12}\bar{z}_{13}})^{\Delta_1-1}(\frac{z_{51}}{z_{12}\bar{z}_{23}})(\frac{z_{51}\bar{z}_{35}}{z_{12}\bar{z}_{23}})^{\Delta_2-1}(\frac{z_{25}\bar{z}_{51}}{z_{23}\bar{z}_{13}})^{\Delta_3-1}\frac{\bar{z}_{12}\bar{z}_{23}}{z_{35}z_{15}}\delta(x-\bar{x})\\
&+\frac{1}{2}\pi\delta(\sum_i\lambda_i)(\Delta_3-1)(\frac{z_{25}}{z_{23}\bar{z}_{13}})(\frac{z_{25}\bar{z}_{35}}{z_{12}\bar{z}_{13}})^{\Delta_1-1}(\frac{z_{51}\bar{z}_{35}}{z_{12}\bar{z}_{23}})^{\Delta_2}(\frac{z_{25}\bar{z}_{51}}{z_{23}\bar{z}_{13}})^{\Delta_3-2}\frac{\bar{z}_{12}\bar{z}_{23}}{z_{35}z_{15}}\delta(x-\bar{x})\\
&+\frac{1}{2}\pi\delta(\sum_i\lambda_i)(\frac{z_{25}\bar{z}_{35}}{z_{12}\bar{z}_{13}})^{\Delta_1-1}(\frac{z_{51}\bar{z}_{35}}{z_{12}\bar{z}_{23}})^{\Delta_2}(\frac{z_{25}\bar{z}_{51}}{z_{23}\bar{z}_{13}})^{\Delta_3-1}\frac{\bar{z}_{12}\bar{z}_{23}}{z_{35}z_{15}}\bar{\partial}_5\delta(x-\bar{x}).
\fe
Then we write down the $\mathcal{O}(\frac{\bar{z}_{45}}{z_{45}})$ order terms of $\ma{A}_{2\rightarrow3}$.
\ie
\label{fivepointh}
\ma{A}_{2\rightarrow3}|_{\mathcal{O}(\frac{\bar{z}_{45}}{z_{45}})}&=\frac{\pi\bar{z}_{45}}{4z_{45}}\delta(\sum_{i=1}^5\lambda_i)B(\Delta_4,\Delta_5-1)\prod_i^3(\epsilon_i\sigma_{i,1})^{\Delta_i-1}
\Big[\frac{\sigma_{1,3}}{\sigma_{1,1}}(\Delta_1-1)\epsilon_2\sigma_{2,1}+\sigma_{2,3}\epsilon_2\Delta_2\\&+\frac{\sigma_{3,3}}{\sigma_{3,1}}(\Delta_3-1)\epsilon_2\sigma_{2,1}\Big]\Big(\frac{|z_{12}|^2|z_{23}|^2}{z_{12}z_{23}z_{34}z_{51}}\Big)\delta(x-\bar{x})\\
&-\frac{\pi\bar{z}_{45}}{4z_{45}}\delta(\sum_{i=1}^5\lambda_i)B(\Delta_4,\Delta_5-1)
\Big(\frac{z_{25}\bar{z}_{35}}{z_{12}\bar{z}_{13}}\Big)^{\Delta_1-1}\Big(\frac{z_{51}\bar{z}_{35}}{z_{12}\bar{z}_{23}}\Big)^{\Delta_2}\Big(\frac{z_{25}\bar{z}_{51}}{z_{23}\bar{z}_{13}}\Big)^{\Delta_3-1}\Big(\frac{\bar{z}_{12}\bar{z}_{23}}{z_{35}z_{51}}\Big)\\
&\times(\frac{x}{\bar{z}_{25}}\partial_x+\frac{\bar{x}}{\bar{z}_{35}}\partial_{\bar{x}})\delta(x-\bar{x})
\fe 
We find that
\ie
\label{fivepointh2}
\ma{A}_{2\rightarrow3}|_{\mathcal{O}(\frac{\bar{z}_{45}}{z_{45}})}&=-\frac{\bar{z}_{45}}{2z_{45}}\delta(\sum_{i=1}^5\lambda_i)B(\Delta_4,\Delta_5-1)\bar{\partial}_{5}\mathcal{A}^{\Delta_5\rightarrow\Delta_4+\Delta_5-1}_{2\rightarrow2}
\fe 
as expected. We can still use \eqref{famp} to obtain the OPE at the order $\ma{O}(\frac{\bar{z}_{45}}{z_{45}})$:
\ie
\ma{O}^a_{\Delta_4}(z_4)\ma{O}^b_{\Delta_5}(z_5)|_{\ma{O}(\frac{\bar{z}_{45}}{z_{45}})}\sim-\frac{f^{abc}B(\Delta_4,\Delta_5-1)\bar{z}_{45}}{2z_{45}}\bar{\partial}\ma{O}^c_{\Delta_4+\Delta_5-1}(z_5).
\fe
Moreover, it is not hard to find that
\ie
\ma{O}^a_{\Delta_4}(z_4)\ma{O}^b_{\Delta_5}(z_5)|_{\ma{O}(\frac{\bar{z}_{45}^n}{z_{45}})}\sim-\frac{f^{abc}B(\Delta_4-1+n,\Delta_5-1)\bar{z}^n_{45}}{2n!z_{45}}\bar{\partial}^n\ma{O}^c_{\Delta_4+\Delta_5-1}(z_5).
\fe

\subsection{Terms at the order $\ma{O}(1)$}
Now we turn to the $\ma{O}(1)$ order terms.
We have 
\ie
\label{fivepoint!!!}
\ma{A}_{2\rightarrow3}|_{\mathcal{O}(1)}&=\frac{\pi}{4}\delta(\sum_{i=1}^5\lambda_i)B(\Delta_4,\Delta_5-1)\prod_i^3(\epsilon_i\sigma_{i,1})^{\Delta_i-1}
\Big[\frac{\sigma_{1,2}}{\sigma_{1,1}}(\Delta_1-1)\epsilon_2\sigma_{2,1}+\sigma_{2,2}\epsilon_2\Delta_2\\&+\frac{\sigma_{3,2}}{\sigma_{3,1}}(\Delta_3-1)\epsilon_2\sigma_{2,1}\Big]\Big(\frac{|z_{12}|^2|z_{23}|^2}{z_{12}z_{23}z_{35}z_{51}}\Big)\delta(x-\bar{x})\\
&+\frac{\pi}{4}\delta(\sum_{i=1}^5\lambda_i)B(\Delta_4,\Delta_5-1)
\Big(\frac{z_{25}\bar{z}_{35}}{z_{12}\bar{z}_{13}}\Big)^{\Delta_1-1}\Big(\frac{z_{51}\bar{z}_{35}}{z_{12}\bar{z}_{23}}\Big)^{\Delta_2}\Big(\frac{z_{25}\bar{z}_{51}}{z_{23}\bar{z}_{13}}\Big)^{\Delta_3-1}\Big(\frac{\bar{z}_{12}\bar{z}_{23}}{z_{35}z_{51}}\Big)\frac{1}{z_{35}}\delta(x-\bar{x})\\
&+\frac{\pi}{4}\delta(\sum_{i=1}^5\lambda_i)(B(\Delta_4,\Delta_5-1)-B(\Delta_4+1,\Delta_5-1))\Big(\frac{z_{25}\bar{z}_{35}}{z_{12}\bar{z}_{13}}\Big)^{\Delta_1-1}\Big(\frac{z_{51}\bar{z}_{35}}{z_{12}\bar{z}_{23}}\Big)^{\Delta_2}\Big(\frac{z_{25}\bar{z}_{51}}{z_{23}\bar{z}_{13}}\Big)^{\Delta_3-1}\\&\times\Big(\frac{\bar{z}_{12}\bar{z}_{23}}{z_{35}z_{51}}\Big)
\delta(x-\bar{x})(\frac{1}{z_{15}}-\frac{1}{z_{35}}).
\fe 
In this case, we find that 
\ie 
\mathcal{A}_{2\rightarrow3}|_{\mathcal{O}(1)}=&-\frac{1}{2}B(\Delta_4,\Delta_5-1)\partial_5\ma{A}^{\Delta_5\rightarrow\Delta_4+\Delta_5-1}_{2\rightarrow2}+\frac{1}{2z_{35}}B(\Delta_4,\Delta_5-1)\ma{A}^{\Delta_5\rightarrow\Delta_4+\Delta_5-1}_{2\rightarrow2}\\
&-\frac{z_{13}}{2z_{15}z_{35}}B(\Delta_4+1,\Delta_5-1)\ma{A}^{\Delta_5\rightarrow\Delta_4+\Delta_5-1}_{2\rightarrow2}.
\fe 
It is not hard to find that the first term comes from the $SL(2,\mathbb{C})$ descendant, and the other terms come from $S$-algebra descendants. One can also obtain the full OPE using \eqref{famp}. In this case, color-ordered amplitudes like $\ma{A}(12435)$ will also contribute, which makes the calculation very complicated. We will not write down this OPE explicitly in this paper and we expect a more elegant form which is a combination of the null states \cite{Banerjee:2023jne}.

\section{Outlooks}
In this paper, we developed a new formula \eqref{ca} (with the special choice of the reference spinor) for celestial color-ordered SDYM. This formula reveals the leading OPE limit of the celestial amplitude explicitly and also allows us to compute the celestial amplitudes from lower-point celestial BG currents. Hence we can also regard this formula as a recursion formula in some sense. A quite interesting question is whether we can express the (celestial) self-dual gravity amplitudes in a similar formalism to ours. In gravity theory, there is no canonical ordering and we cannot use region momentum, which makes it difficult to find such a formula. If there exists such a formula, it may help us to understand the structure of the (celestial) self-dual gravity theory such as some properties given by \cite{Banerjee:2023bni}.

We also gave some results of the OPE limit in this paper. However, we have not organized them as a combination of the null states as suggested in \cite{Banerjee:2023jne}. It will be interesting to explore if the conformal dimensions of null states of all orders of OPE have a lower bound. This problem is similar to the statement for celestial self-dual gravity in \cite{Banerjee:2023bni} where the authors give some evidence in the first few orders of OPE.

Another interesting direction is the duality between the sourced SDYM and the Liouvelle CFT \cite{Melton:2023lnz}. In sourced SDYM, there exist high-point tree amplitudes due to the scalar source. It will be interesting to generalize their result to the loop level and figure out what role our formula can play in this generalization.

\section*{Acknowledgement}
YT is partly supported by the National Key R\&D Program of China (NO. 2020YFA0713000).

\appendix
\section{Massive spinors on the celestial sphere and the effective propagator}\label{app1}
In this appendix, we will show how to translate a massive momentum to massive spinors and then write down the explicit form of the effective propagator. First, we will review our conventions for massless spinors. For massless spinors, we use the usual ones \cite{Pasterski:2017ylz,Elvang:2013cua}:
\ie
|k\rangle=\epsilon\sqrt{2\omega}(1,z)^{T},\langle k|=\epsilon\sqrt{2\omega}(-z,1),|k]=\sqrt{2\omega}(-\bar{z},1)^{T},[k|=\sqrt{2\omega}(1,\bar{z})
\fe
and
\ie
k_{\alpha\dot{\alpha}}=k_\mu(\sigma^{\mu})_{\alpha\dot{\alpha}}=-|k]_\alpha\langle k|_{\dot{\alpha}},k^{\dot{\alpha}\alpha}=k_\mu(\bar{\sigma}^{\mu})^{\dot{\alpha}\alpha}=-|k\rangle^{\dot{\alpha}}[k|^\alpha
\fe
where $\sigma^\mu=(1,\sigma^i)$ and $\bar{\sigma}^{\mu}=(1,-\sigma^i)$ with $\sigma^i$ the Pauli matrices. Spinor indices are raised(lowered) using
\ie
\epsilon^{12}=-\epsilon^{21}=-\epsilon_{12}=\epsilon_{21}=1
\fe

For a massive momentum, we can also derive the spinor-helicity formalism \cite{Arkani-Hamed:2017jhn}. Another good review is \cite{Chiodaroli:2021eug}. In the celestial case, things are more transparent. First we can divide the massive momentum $k^\mu$ into two massless vectors $\hat{q}^\mu$ and $a^\mu=(1,0,0,-1)$:
\ie
k^{\mu}=\frac{m}{2y}\epsilon\hat{q}^{\mu}+\frac{my}{2}\epsilon(1,0,0,-1).
\fe
Then
\ie
k^{\dot{\alpha}\alpha}=-\frac{m}{2y}\epsilon|\hat{q}\rangle[\hat{q}|-\frac{my}{2}\epsilon|a\rangle[a|=-|k^a\rangle[k_a|
\fe
where
\ie
\ [a|=(0,\sqrt{2}),|a\rangle=(0,\sqrt{2})^T
\fe
and $a$ is a index running over the $\hat{q}^{\mu}$ spinor part and the $a^{\mu}$ spinor part. We also have
\ie
k_{\alpha\dot{\alpha}}=-\frac{m}{2y}\epsilon|\hat{q}]\langle\hat{q}|-\frac{my}{2}\epsilon|a]\langle a|=-|k^a]\langle k_a|
\fe
where
\ie
|a]=(-\sqrt{2},0)^T,\langle a|=(-\sqrt{2},0).
\fe
Then we can write down the spinors for the massive momentum $k^{\mu}$
\ie
&|k^a\rangle=(\sqrt{\frac{my}{2}}|a\rangle,\sqrt{\frac{m}{2y}}|\hat{q}\rangle)^{T}=\epsilon((0,\sqrt{my}),\sqrt{\frac{m}{y}}(1,z))^{T},\\
&[k_a|=((0,\sqrt{my}),\sqrt{\frac{m}{y}}(1,\bar{z})),\\
&\langle k_a|=\epsilon(\sqrt{\frac{m}{y}}(-z,1),(-\sqrt{my},0)),\\
&|k^a]=(\sqrt{\frac{m}{y}}(-\bar{z},1),(-\sqrt{my},0))^{T}.
\fe

For the special case $|q\rangle=(0,1)$,$\langle q|=(-1,0)$, we have
\ie
\langle qk_i\rangle[k_ik_R^b]\langle k_{Rb}q\rangle=2\epsilon_i\epsilon_R\omega_i\frac{m_R}{y_R}\bar{z}_{Ri}
\fe
for a massless momentum $k_i$. Then the effective propagator is
\ie
(\sum_{k=n}^i\langle qk_k\rangle[k_kk_R^b]\langle k_{Rb}q\rangle)^2=(2\sum_{k=n}^i\epsilon_k\epsilon_R\omega_i\frac{m_R}{y_R}\bar{z}_{Rk})^2=4\frac{m_R^2}{y_R^2}\sum_{k,l=1}^{i}\epsilon_k\epsilon_l\omega_k\omega_l\bar{z}_{kR}\bar{z}_{lR}.
\fe
For another special case $|q\rangle=(1,z_3),\langle q|=(-z_3,1)$ (i.e. $q\propto k_3$), we have
\ie
\langle qk_i\rangle[k_ik_R^b]\langle k_{Rb}q\rangle=2z_{3i}(\epsilon_i\epsilon_R\omega_i\frac{m_R}{y_R}\bar{z}_{Ri}z_{3R}-\epsilon_i\epsilon_R\omega_im_Ry_R)
\fe
and
\ie
4\frac{m_R^2}{y^2_R}\sum_{k,l=n}^i\omega_k\omega_l\Pi_{k,l}(k_R):=&(\sum_{k=n}^i\langle qk_k\rangle[k_kk_R^b]\langle k_{Rb}q\rangle)^2\\
=&4\frac{m_R^2}{y_R^2}\sum_{k,l=n}^i\omega_k\omega_lz_{3k}z_{3l}(\epsilon_k\bar{z}_{Rk}z_{3R}-\epsilon_ky^2_R)(\epsilon_l\bar{z}_{Rl}z_{3R}-\epsilon_ly^2_R).
\fe
We will use the latter choice to compute the 4-pt celestial amplitudes to simplify the calculation.

\section{$s$-channel factorization of the 4-pt delta function}
In this appendix, we will show that the $s$-channel factorization of the $2\rightarrow2$ delta function leads to the cross-ratio delta function to show that this factorization will give the correct support. In the following derivation, we will ignore the integrands and the Jacobians and only focus on the delta functions. Note that we have integrated $\omega_2$ and $\omega_4$ using the parameterization in \cite{Raclariu:2021zjz}:
\ie
&\delta^{(4)}\sim\int da_Rdy_Rd^2z_R\delta(y_R-\frac{2m_R\omega_3|z_{34}|^2}{m_R^2+4\omega_3^2|z_{34}|^2})\delta^{(2)}(z_R-\frac{m_R^2z_4+4\omega_3^2z_1|z_{34}|^2}{m_R^2+4\omega_3^2|z_{34}|^2})\\
&\times\delta(y_R-\frac{2m_R\omega_1|z_{12}|^2}{m_R^2+4\omega_1^2|z_{12}|^2})\delta^{(2)}(z_R-\frac{m_R^2z_2+4\omega_1^2z_1|z_{12}|^2}{m_R^2+4\omega_1^2|z_{12}|^2})\\
&=\int da_Rdy_R\delta(y_R-\frac{2m_R\omega_3|z_{34}|^2}{m_R^2+4\omega_3^2|z_{34}|^2})\delta(y_R-\frac{2m_R\omega_1|z_{12}|^2}{m_R^2+4\omega_1^2|z_{12}|^2})\\
&\times\delta^{(2)}(\frac{m_R^2z_4+4\omega_3^2z_3|z_{34}|^2}{m_R^2+4\omega_3^2|z_{34}|^2}-\frac{m_R^2z_2+4\omega_1^2z_1|z_{12}|^2}{m_R^2+4\omega_1^2|z_{12}|^2}).
\fe
Note that $m_R$ is not an independent variable and we have $m_R=2a_Ry_R$. Thus the delta function should be
\ie
\delta(y_R-\frac{2m_R\omega_3|z_{34}|^2}{m_R^2+4\omega_3^2|z_{34}|^2})=\delta(y_R-\frac{a_Ry_R\omega_3|z_{34}|^2}{a_R^2y_R^2+\omega_3^2|z_{34}|^2}).
\fe
We do not consider the support $y_R=0$ as it corresponds to $k_R$ on-shell. Then we have
\ie
\delta(y_R-\frac{a_Ry_R\omega_3|z_{34}|^2}{a_R^2y_R^2+\omega_3^2|z_{34}|^2})=\frac{\omega_3|z_{34}|^2}{2a_Ry^2_R}\delta(y_R-\frac{\sqrt{-\omega_3^2|z_{34}|^2+a_R\omega_3|z_{34}|^2}}{a_R}).
\fe
Another useful formula from this delta function is
\ie
m_R^2+4\omega_3^2|z_{34}|^2=4a_R\omega_3|z_{34}|^2
\fe
and the other $y_R$ delta function will become a delta function of $a_R$. We can do the $y_R$ integral first, then the delta functions will be
\ie
&\delta^{(4)}\sim\int da_R\delta(\frac{\sqrt{-\omega_3^2|z_{34}|^2+a_R\omega_3|z_{34}|^2}}{a_R}-\frac{\sqrt{-\omega_1^2|z_{12}|^2+a_R\omega_1|z_{12}|^2}}{a_R})\\
&\times\delta^{(2)}(\frac{(a_R-\omega_3)z_4+\omega_3z_3}{a_R}-\frac{(a_R-\omega_1)z_2+\omega_1z_1}{a_R}).
\fe
Note that
\ie
&\delta(\frac{\sqrt{-\omega_3^2|z_{34}|^2+a_R\omega_3|z_{34}|^2}}{a_R}-\frac{\sqrt{-\omega_1^2|z_{12}|^2+a_R\omega_1|z_{12}|^2}}{a_R})\sim\delta(a_R-\frac{\omega_3^2|z_{34}|^2-\omega_1^2|z_{12}|^2}{\omega_3|z_{34}|^2-\omega_1|z_{12}|^2}).
\fe
Then we can do the $a_R$ integral:
\ie
\delta^{(4)}\sim\delta^{(2)}(\omega_1^2|z_{12}|^2z_{14}+\omega_3^2|z_{34}|^2z_{32}+\omega_1\omega_3|z_{34}|^2z_{21}+\omega_1\omega_3|z_{12}|^2z_{43}).
\fe
Using the momentum conservation, we have $\omega_1=\frac{\bar{z}_{34}z_{23}}{\bar{z}_{41}z_{12}}\omega_3$. Then we have
\ie
&(\frac{\bar{z}_{34}z_{23}}{\bar{z}_{41}z_{12}})^2|z_{12}|^2z_{14}+|z_{34}|^2z_{32}+\frac{\bar{z}_{34}z_{23}}{\bar{z}_{41}z_{12}}|z_{34}|^2z_{21}+\frac{\bar{z}_{34}z_{23}}{\bar{z}_{41}z_{12}}|z_{12}|^2z_{43}=0\\
&\rightarrow\bar{z}^2_{34}z_{23}\bar{z}_{12}z_{14}-\bar{z}^2_{41}z_{12}|z_{34}|^2-\bar{z}_{41}z_{12}\bar{z}_{34}|z_{34}|^2+\bar{z}_{41}\bar{z}_{34}|z_{12}|^2z_{43}=0\\
&\rightarrow\bar{z}_{34}z_{23}\bar{z}_{12}z_{14}+\bar{z}_{41}z_{12}\bar{z}_{13}z_{34}+\bar{z}_{41}|z_{12}|^2z_{43}=0\\
&\rightarrow\bar{z}_{34}z_{23}\bar{z}_{12}z_{14}-\bar{z}_{14}z_{12}\bar{z}_{23}z_{34}=0
\fe
which is exactly the cross-ratio delta function. We can now use the Schouten identity to match it with the usual cross-ratio delta function. The Schouten identity gives:
\ie
&z_{12}z_{34}+z_{13}z_{42}+z_{14}z_{23}=0,\\
&\bar{z}_{12}\bar{z}_{34}+\bar{z}_{13}\bar{z}_{42}+\bar{z}_{14}\bar{z}_{23}=0.
\fe
Then
\ie
&\bar{z}_{34}z_{23}\bar{z}_{12}z_{14}-\bar{z}_{14}z_{12}\bar{z}_{23}z_{34}=0\rightarrow\bar{z}_{12}\bar{z}_{34}(z_{12}z_{34}+z_{13}z_{42})-z_{12}z_{34}(\bar{z}_{12}\bar{z}_{34}+\bar{z}_{13}\bar{z}_{42})=0\\
&\rightarrow\bar{z}_{12}\bar{z}_{34}z_{13}z_{24}-z_{12}z_{34}\bar{z}_{13}\bar{z}_{24}=0.
\fe
Finally we can write down the delta function:
\ie
&\delta^{(2)}(\omega_1^2|z_{12}|^2z_{14}+\omega_3^2|z_{34}|^2z_{32}+\omega_1\omega_3|z_{34}|^2z_{21}+\omega_1\omega_3|z_{12}|^2z_{43})\\
&\sim\delta(\omega_1-\frac{\bar{z}_{34}z_{23}}{\bar{z}_{41}z_{12}}\omega_3)\delta(\bar{z}_{12}\bar{z}_{34}z_{13}z_{24}-z_{12}z_{34}\bar{z}_{13}\bar{z}_{24}).
\fe
Thus we obtain the cross-ratio delta function from $\delta^{(4)}$.

\bibliographystyle{JHEP}
\bibliography{CSDYM}

\providecommand{\href}[2]{#2}\begingroup\raggedright\begin{thebibliography}{10}

\bibitem{Maldacena:1997re}
J.M.~Maldacena, \emph{{The Large N limit of superconformal field theories and
  supergravity}}, \href{https://doi.org/10.4310/ATMP.1998.v2.n2.a1}{\emph{Adv.
  Theor. Math. Phys.} {\bfseries 2} (1998) 231}
  [\href{https://arxiv.org/abs/hep-th/9711200}{{\ttfamily hep-th/9711200}}].

\bibitem{Pasterski:2017kqt}
S.~Pasterski and S.-H.~Shao, \emph{{Conformal basis for flat space
  amplitudes}}, \href{https://doi.org/10.1103/PhysRevD.96.065022}{\emph{Phys.
  Rev. D} {\bfseries 96} (2017) 065022}
  [\href{https://arxiv.org/abs/1705.01027}{{\ttfamily 1705.01027}}].

\bibitem{Raclariu:2021zjz}
A.-M.~Raclariu, \emph{{Lectures on Celestial Holography}},
  \href{https://arxiv.org/abs/2107.02075}{{\ttfamily 2107.02075}}.

\bibitem{Pasterski:2021rjz}
S.~Pasterski, \emph{{Lectures on celestial amplitudes}},
  \href{https://doi.org/10.1140/epjc/s10052-021-09846-7}{\emph{Eur. Phys. J. C}
  {\bfseries 81} (2021) 1062}
  [\href{https://arxiv.org/abs/2108.04801}{{\ttfamily 2108.04801}}].

\bibitem{Chang:2022jut}
C.-M.~Chang, W.~Cui, W.-J.~Ma, H.~Shu and H.~Zou, \emph{{Shadow celestial
  amplitudes}}, \href{https://doi.org/10.1007/JHEP02(2023)017}{\emph{JHEP}
  {\bfseries 02} (2023) 017}
  [\href{https://arxiv.org/abs/2210.04725}{{\ttfamily 2210.04725}}].

\bibitem{Freidel:2022skz}
L.~Freidel, D.~Pranzetti and A.-M.~Raclariu, \emph{{A discrete basis for
  celestial holography}},  \href{https://arxiv.org/abs/2212.12469}{{\ttfamily
  2212.12469}}.

\bibitem{Cotler:2023qwh}
J.~Cotler, N.~Miller and A.~Strominger, \emph{{An integer basis for celestial
  amplitudes}}, \href{https://doi.org/10.1007/JHEP08(2023)192}{\emph{JHEP}
  {\bfseries 08} (2023) 192}
  [\href{https://arxiv.org/abs/2302.04905}{{\ttfamily 2302.04905}}].

\bibitem{Hu:2022bpa}
Y.~Hu and S.~Pasterski, \emph{{Celestial recursion}},
  \href{https://doi.org/10.1007/JHEP01(2023)151}{\emph{JHEP} {\bfseries 01}
  (2023) 151} [\href{https://arxiv.org/abs/2208.11635}{{\ttfamily
  2208.11635}}].

\bibitem{Tao:2023wls}
Y.-X.~Tao, \emph{{Celestial Berends-Giele current}},
  \href{https://doi.org/10.1007/JHEP09(2023)193}{\emph{JHEP} {\bfseries 09}
  (2023) 193} [\href{https://arxiv.org/abs/2307.14772}{{\ttfamily
  2307.14772}}].

\bibitem{Chang:2023ttm}
C.-M.~Chang, R.~Liu and W.-J.~Ma, \emph{{Split representation in celestial
  holography}},  \href{https://arxiv.org/abs/2311.08736}{{\ttfamily
  2311.08736}}.

\bibitem{Melton:2023bjw}
W.~Melton, A.~Sharma and A.~Strominger, \emph{{Celestial Leaf Amplitudes}},
  \href{https://arxiv.org/abs/2312.07820}{{\ttfamily 2312.07820}}.

\bibitem{Jiang:2022hho}
H.~Jiang, \emph{{Celestial Mellin amplitude}},
  \href{https://doi.org/10.1007/JHEP10(2022)042}{\emph{JHEP} {\bfseries 10}
  (2022) 042} [\href{https://arxiv.org/abs/2208.01576}{{\ttfamily
  2208.01576}}].

\bibitem{Strominger:2021lvk}
A.~Strominger, \emph{{w(1+infinity) and the Celestial Sphere}},
  \href{https://arxiv.org/abs/2105.14346}{{\ttfamily 2105.14346}}.

\bibitem{Costello:2022jpg}
K.~Costello, N.M.~Paquette and A.~Sharma, \emph{{Top-Down Holography in an
  Asymptotically Flat Spacetime}},
  \href{https://doi.org/10.1103/PhysRevLett.130.061602}{\emph{Phys. Rev. Lett.}
  {\bfseries 130} (2023) 061602}
  [\href{https://arxiv.org/abs/2208.14233}{{\ttfamily 2208.14233}}].

\bibitem{Costello:2023hmi}
K.~Costello, N.M.~Paquette and A.~Sharma, \emph{{Burns space and holography}},
  \href{https://doi.org/10.1007/JHEP10(2023)174}{\emph{JHEP} {\bfseries 10}
  (2023) 174} [\href{https://arxiv.org/abs/2306.00940}{{\ttfamily
  2306.00940}}].

\bibitem{Stieberger:2022zyk}
S.~Stieberger, T.R.~Taylor and B.~Zhu, \emph{{Celestial Liouville theory for
  Yang-Mills amplitudes}},
  \href{https://doi.org/10.1016/j.physletb.2022.137588}{\emph{Phys. Lett. B}
  {\bfseries 836} (2023) 137588}
  [\href{https://arxiv.org/abs/2209.02724}{{\ttfamily 2209.02724}}].

\bibitem{Stieberger:2023fju}
S.~Stieberger, T.R.~Taylor and B.~Zhu, \emph{{Yang-Mills as a Liouville
  theory}}, \href{https://doi.org/10.1016/j.physletb.2023.138229}{\emph{Phys.
  Lett. B} {\bfseries 846} (2023) 138229}
  [\href{https://arxiv.org/abs/2308.09741}{{\ttfamily 2308.09741}}].

\bibitem{Banerjee:2023bni}
S.~Banerjee, R.~Mandal, S.~Misra, S.~Panda and P.~Paul, \emph{{All $S$
  invariant gluon OPEs on the celestial sphere}},
  \href{https://arxiv.org/abs/2311.16796}{{\ttfamily 2311.16796}}.

\bibitem{Banerjee:2023zip}
S.~Banerjee, H.~Kulkarni and P.~Paul, \emph{{An infinite family of
  w$_{1+\infty}$ invariant theories on the celestial sphere}},
  \href{https://doi.org/10.1007/JHEP05(2023)063}{\emph{JHEP} {\bfseries 05}
  (2023) 063} [\href{https://arxiv.org/abs/2301.13225}{{\ttfamily
  2301.13225}}].

\bibitem{Banerjee:2023jne}
S.~Banerjee, H.~Kulkarni and P.~Paul, \emph{{Celestial OPE in Self Dual
  Gravity}},  \href{https://arxiv.org/abs/2311.06485}{{\ttfamily 2311.06485}}.

\bibitem{Krasnov:2016emc}
K.~Krasnov, \emph{{Self-Dual Gravity}},
  \href{https://doi.org/10.1088/1361-6382/aa65e5}{\emph{Class. Quant. Grav.}
  {\bfseries 34} (2017) 095001}
  [\href{https://arxiv.org/abs/1610.01457}{{\ttfamily 1610.01457}}].

\bibitem{Albayrak:2020saa}
S.~Albayrak, C.~Chowdhury and S.~Kharel, \emph{{On loop celestial amplitudes
  for gauge theory and gravity}},
  \href{https://doi.org/10.1103/PhysRevD.102.126020}{\emph{Phys. Rev. D}
  {\bfseries 102} (2020) 126020}
  [\href{https://arxiv.org/abs/2007.09338}{{\ttfamily 2007.09338}}].

\bibitem{Fan:2019emx}
W.~Fan, A.~Fotopoulos and T.R.~Taylor, \emph{{Soft Limits of Yang-Mills
  Amplitudes and Conformal Correlators}},
  \href{https://doi.org/10.1007/JHEP05(2019)121}{\emph{JHEP} {\bfseries 05}
  (2019) 121} [\href{https://arxiv.org/abs/1903.01676}{{\ttfamily
  1903.01676}}].

\bibitem{Pasterski:2017ylz}
S.~Pasterski, S.-H.~Shao and A.~Strominger, \emph{{Gluon Amplitudes as 2d
  Conformal Correlators}},
  \href{https://doi.org/10.1103/PhysRevD.96.085006}{\emph{Phys. Rev. D}
  {\bfseries 96} (2017) 085006}
  [\href{https://arxiv.org/abs/1706.03917}{{\ttfamily 1706.03917}}].

\bibitem{Chalmers:1996rq}
G.~Chalmers and W.~Siegel, \emph{{The Selfdual sector of QCD amplitudes}},
  \href{https://doi.org/10.1103/PhysRevD.54.7628}{\emph{Phys. Rev. D}
  {\bfseries 54} (1996) 7628}
  [\href{https://arxiv.org/abs/hep-th/9606061}{{\ttfamily hep-th/9606061}}].

\bibitem{Chattopadhyay:2021udc}
P.~Chattopadhyay and K.~Krasnov, \emph{{One-loop same helicity YM amplitudes
  from BG currents}},
  \href{https://doi.org/10.1007/JHEP03(2022)191}{\emph{JHEP} {\bfseries 03}
  (2022) 191} [\href{https://arxiv.org/abs/2110.00331}{{\ttfamily
  2110.00331}}].

\bibitem{Monteiro:2022nqt}
R.~Monteiro, R.~Stark-Much\~ao and S.~Wikeley, \emph{{Anomaly and double copy
  in quantum self-dual Yang-Mills and gravity}},
  \href{https://doi.org/10.1007/JHEP09(2023)030}{\emph{JHEP} {\bfseries 09}
  (2023) 030} [\href{https://arxiv.org/abs/2211.12407}{{\ttfamily
  2211.12407}}].

\bibitem{Pate:2019lpp}
M.~Pate, A.-M.~Raclariu, A.~Strominger and E.Y.~Yuan, \emph{{Celestial operator
  products of gluons and gravitons}},
  \href{https://doi.org/10.1142/S0129055X21400031}{\emph{Rev. Math. Phys.}
  {\bfseries 33} (2021) 2140003}
  [\href{https://arxiv.org/abs/1910.07424}{{\ttfamily 1910.07424}}].

\bibitem{Melton:2023lnz}
W.~Melton and S.A.~Narayanan, \emph{{Celestial Gluon Amplitudes from the
  Outside In}},  \href{https://arxiv.org/abs/2312.12394}{{\ttfamily
  2312.12394}}.

\bibitem{Elvang:2013cua}
H.~Elvang and Y.-t.~Huang, \emph{{Scattering Amplitudes}},
  \href{https://arxiv.org/abs/1308.1697}{{\ttfamily 1308.1697}}.

\bibitem{Arkani-Hamed:2017jhn}
N.~Arkani-Hamed, T.-C.~Huang and Y.-t.~Huang, \emph{{Scattering amplitudes for
  all masses and spins}},
  \href{https://doi.org/10.1007/JHEP11(2021)070}{\emph{JHEP} {\bfseries 11}
  (2021) 070} [\href{https://arxiv.org/abs/1709.04891}{{\ttfamily
  1709.04891}}].

\bibitem{Chiodaroli:2021eug}
M.~Chiodaroli, H.~Johansson and P.~Pichini, \emph{{Compton black-hole
  scattering for s \ensuremath{\leq} 5/2}},
  \href{https://doi.org/10.1007/JHEP02(2022)156}{\emph{JHEP} {\bfseries 02}
  (2022) 156} [\href{https://arxiv.org/abs/2107.14779}{{\ttfamily
  2107.14779}}].

\end{thebibliography}\endgroup

\end{fmffile}
\end{document}